\documentclass[twoside]{ilcws08}
\usepackage[latin1]{inputenc}
\usepackage[dvips]{graphicx,epsfig,color}
\usepackage{wrapfig,rotating}
\usepackage{amssymb,amsmath,array}

\begin{document}
\def\leqsim{\mathbin{\;\raise1pt\hbox{$<$}\kern-8pt\lower3pt\hbox{$\sim$}\;}}
\def\geqsim{\mathbin{\;\raise1pt\hbox{$>$}\kern-8pt\lower3pt\hbox{$\sim$}\;}}
\renewcommand\topfraction{1.}
\renewcommand\bottomfraction{1.}
\renewcommand\floatpagefraction{0.}
\renewcommand\textfraction{0.}
% Charginos and Neutralinos :
\def\MXN#1{\mbox{$ M_{\tilde{\chi}^0_#1}                                $}}
\def\MXC#1{\mbox{$ M_{\tilde{\chi}^{\pm}_#1}                            $}}
\def\XP#1{\mbox{$ \tilde{\chi}^+_#1                                     $}}
\def\XM#1{\mbox{$ \tilde{\chi}^-_#1                                     $}}
\def\XPM#1{\mbox{$ \tilde{\chi}^{\pm}_#1                                $}}
\def\XN#1{\mbox{$ \tilde{\chi}^0_#1                                     $}}
\def\XNN#1#2{\mbox{$ \tilde{\chi}^0_{#1,#2}                             $}}
\def\MXn{\mbox{$ M_{\tilde{\chi}^0}                                     $}}
\def\MXc{\mbox{$ M_{\tilde{\chi}^{\pm}}                                 $}}
\def\Xp{\mbox{$ \tilde{\chi}^+                                          $}}
\def\Xm{\mbox{$ \tilde{\chi}^-                                          $}}
\def\Xpm{\mbox{$ \tilde{\chi}^{\pm}                                     $}}
\def\Xn{\mbox{$ \tilde{\chi}^0                                          $}}
\def\Xnn{\mbox{$ \tilde{\chi}^0                                         $}}
\def\p#1{\mbox{$ \mbox{\bf p}_1                                         $}}
\def\Xgen{\mbox{$ \tilde{\chi}                                          $}}
\def\Mlsp{\mbox{$ M_{\mathrm {LSP}}                                     $}}
\def\lsp{\mbox{$ {\mathrm {LSP}}                                     $}}
\newcommand{\Ptmis}   {\mbox{$/\mkern-11mu P_t \,                          $}}
\newcommand{\Tpmis}   {\mbox{$\theta_{/\mkern-11mu p}                      $}}
% sparticles
\newcommand{\grav}    {\mbox{$ \tilde{\mathrm G}                           $}}
\newcommand{\Gino}    {\mbox{$ \tilde{\mathrm G}                           $}}
\newcommand{\tanb}    {\mbox{$ \tan \beta                                  $}}
\newcommand{\smu}     {\mbox{$ \tilde{\mu}                                 $}}
\newcommand{\smur}    {\mbox{$ \tilde{\mu}_{\mathrm R}                     $}}
\newcommand{\smul}    {\mbox{$ \tilde{\mu}_{\mathrm L}                     $}}
\newcommand{\msmu}    {\mbox{$ M_{\tilde{\mu}}                             $}}
\newcommand{\msmur}   {\mbox{$ M_{\tilde{\mu}_{\mathrm R}}                 $}}
\newcommand{\msmul}   {\mbox{$ M_{\tilde{\mu}_{\mathrm L}}                 $}}
\newcommand{\sel}     {\mbox{$ \tilde{\mathrm e}                           $}}
\newcommand{\sell}    {\mbox{$ \tilde{\mathrm e}_{\mathrm L}               $}}
\newcommand{\selr}    {\mbox{$ \tilde{\mathrm e}_{\mathrm R}               $}}
\newcommand{\msel}    {\mbox{$ M_{\tilde{\mathrm e}}                       $}}
\newcommand{\snu}     {\mbox{$ \tilde\nu                                   $}}
\newcommand{\msnu}    {\mbox{$ m_{\tilde\nu}                               $}}
\newcommand{\msell}   {\mbox{$ M_{\tilde{\mathrm e}_{\mathrm L}}           $}}
\newcommand{\mselr}   {\mbox{$ M_{\tilde{\mathrm e}_{\mathrm R}}           $}}
\newcommand{\fe}      {\mbox{$ \mathrm f                                   $}}
\newcommand{\feb}     {\mbox{$ \overline{\mathrm f}                        $}}
\newcommand{\sfe}     {\mbox{$ \tilde{\mathrm f}                           $}}
\newcommand{\sfeb}    {\mbox{$ \overline{\tilde{\mathrm f}}                $}}
\newcommand{\sfel}    {\mbox{$ \tilde{\mathrm f}_{\mathrm L}               $}}
\newcommand{\sfer}    {\mbox{$ \tilde{\mathrm f}_{\mathrm R}               $}}
\newcommand{\sfelb}   {\mbox{$ \overline{\tilde{\mathrm f}_{\mathrm L}}    $}}
\newcommand{\sferb}   {\mbox{$ \overline{\tilde{\mathrm f}_{\mathrm R}}    $}}
\newcommand{\msfe}    {\mbox{$ M_{\tilde{\mathrm f}}                       $}}
\newcommand{\sle}     {\mbox{$ \tilde{\ell}                                $}}
\newcommand{\sq}     {\mbox{$ \tilde{q}                                $}}
\newcommand{\sqr}     {\mbox{$ \tilde{q}_{\mathrm R}                                $}}
\newcommand{\sql}     {\mbox{$ \tilde{q}_{\mathrm L}                                $}}
\newcommand{\msle}    {\mbox{$ M_{\tilde{\ell}}                            $}}
\newcommand{\stau}    {\mbox{$ \tilde{\tau}                                $}}
\newcommand{\stone}   {\mbox{$ \tilde{\tau}_1                              $}}
\newcommand{\sttwo}   {\mbox{$ \tilde{\tau}_2                              $}}
\newcommand{\staur}   {\mbox{$ \tilde{\tau}_{\mathrm R}                    $}}
\newcommand{\mstau}   {\mbox{$ M_{\tilde{\tau}}                            $}}
\newcommand{\mstone}  {\mbox{$ M_{\tilde{\tau}_1}                          $}}
\newcommand{\msttwo}  {\mbox{$ M_{\tilde{\tau}_2}                          $}}
\newcommand{\stq}     {\mbox{$ \tilde {\mathrm t}                          $}}
\newcommand{\stqone}  {\mbox{$ \tilde {\mathrm t}_1                        $}}
\newcommand{\stqtwo}  {\mbox{$ \tilde {\mathrm t}_2                        $}}
\newcommand{\msq}    {\mbox{$ M_{\tilde {\mathrm q}}                      $}}
\newcommand{\mstq}    {\mbox{$ M_{\tilde {\mathrm t}}                      $}}
\newcommand{\sbq}     {\mbox{$ \tilde {\mathrm b}                          $}}
\newcommand{\sbqone}  {\mbox{$ \tilde {\mathrm b}_1                        $}}
\newcommand{\sbqtwo}  {\mbox{$ \tilde {\mathrm b}_2                        $}}
\newcommand{\msbq}    {\mbox{$ M_{\tilde {\mathrm b}}                      $}}
\newcommand{\msbqone}    {\mbox{$ M_{\tilde {\mathrm b}_1}                 $}}
\newcommand{\msbqtwo}    {\mbox{$ M_{\tilde {\mathrm b}_2}                 $}}
\newcommand{\mstqone}    {\mbox{$ M_{\tilde {\mathrm t}_1}                 $}}
\newcommand{\mstqtwo}    {\mbox{$ M_{\tilde {\mathrm t}_2}                 $}}
% bosons
\newcommand{\An}      {\mbox{$ {\, \mathrm A}^0                               $}}
\newcommand{\hn}      {\mbox{$ {\, \mathrm h}^0                               $}}
\newcommand{\Zn}      {\mbox{$ {\, \mathrm Z}                                 $}}
\newcommand{\Zstar}   {\mbox{$ {\, \mathrm Z}^*                               $}}
\newcommand{\Hn}      {\mbox{$ {\, \mathrm H}^0                               $}}
\newcommand{\HP}      {\mbox{$ {\, \mathrm H}^+                               $}}
\newcommand{\HM}      {\mbox{$ {\, \mathrm H}^-                               $}}
\newcommand{\W}      {\mbox{$ {\, \mathrm W}                               $}}
\newcommand{\Wp}      {\mbox{$ {\, \mathrm W}^+                               $}}
\newcommand{\Wm}      {\mbox{$ {\, \mathrm W}^-                               $}}
\newcommand{\Wstar}   {\mbox{$ {\, \mathrm W}^*                               $}}
% bosn pairs
\newcommand{\WW}      {\mbox{$ {\, \mathrm W}^+{\mathrm W}^-                  $}}
\newcommand{\ZZ}      {\mbox{$ {\, \mathrm{Z Z}}$}}
\newcommand{\HZ}      {\mbox{$ {\, \mathrm H}^0 {\mathrm Z}                   $}}
\newcommand{\GW}      {\mbox{$ \Gamma_{\mathrm W}                          $}}
\newcommand{\Zg}      {\mbox{$ \Zn \gamma                                  $}}
\newcommand{\Zorg}      {\mbox{$ \Zn / \gamma                                  $}}
\newcommand{\sqs}     {\mbox{$ \sqrt{s}                                    $}}
\newcommand{\epm}     {\mbox{$ {\, \mathrm e}^{\pm}                           $}}
% fermion pairs
\newcommand{\ee}      {\mbox{$ {\, \mathrm e}^+ {\mathrm e}^-                 $}}
\newcommand{\mumu}    {\mbox{$ \, \mu^+ \mu^-                                 $}}
\newcommand{\tautau}  {\mbox{$ \, \tau^+ \tau^-                               $}}
\newcommand{\eeto}    {\mbox{$ {\, \mathrm e}^+ {\mathrm e}^- \to             $}}
\newcommand{\ellell}  {\mbox{$ \, \ell^+ \ell^-                               $}}
\newcommand{\eeWW}    {\mbox{$ \, \ee \rightarrow \, \WW                         $}}
% units
\newcommand{\MeV}     {\mbox{$ {\mathrm{MeV}}                              $}}
\newcommand{\MeVc}    {\mbox{$ {\mathrm{MeV}}/c                            $}}
\newcommand{\MeVcc}   {\mbox{$ {\mathrm{MeV}}/c^2                          $}}
\newcommand{\GeV}     {\mbox{$ {\mathrm{GeV}}                              $}}
\newcommand{\GeVc}    {\mbox{$ {\mathrm{GeV}}/c                            $}}
\newcommand{\GeVcc}   {\mbox{$ {\mathrm{GeV}}/c^2                          $}}
\newcommand{\TeV}     {\mbox{$ {\mathrm{TeV}}                              $}}
\newcommand{\TeVc}    {\mbox{$ {\mathrm{TeV}}/c                            $}}
\newcommand{\TeVcc}   {\mbox{$ {\mathrm{TeV}}/c^2                          $}}
\newcommand{\pbi}     {\mbox{$ {\mathrm{pb}}^{-1}                          $}}
%  masses
\newcommand{\MZ}      {\mbox{$ m_{{\mathrm Z}}                             $}}
\newcommand{\MW}      {\mbox{$ m_{\mathrm W}                               $}}
\newcommand{\MA}      {\mbox{$ m_{\mathrm A}                               $}}
\newcommand{\GF}      {\mbox{$ {\mathrm G}_{\mathrm F}                     $}}
\newcommand{\MH}      {\mbox{$ m_{{\mathrm H}^0}                           $}}
\newcommand{\MHP}     {\mbox{$ m_{{\mathrm H}^\pm}                         $}}
\newcommand{\MSH}     {\mbox{$ m_{{\mathrm h}^0}                           $}}
\newcommand{\MT}      {\mbox{$ m_{\mathrm t}                               $}}
\newcommand{\GZ}      {\mbox{$ \Gamma_{{\mathrm Z} }                       $}}

\newcommand{\TT}      {\mbox{$ \mathrm T                                   $}}
\newcommand{\UU}      {\mbox{$ \mathrm U                                   $}}
\newcommand{\alphmz}  {\mbox{$ \alpha (m_{{\mathrm Z}})                    $}}
\newcommand{\alphas}  {\mbox{$ \alpha_{\mathrm s}                          $}}
\newcommand{\alphmsb} {\mbox{$ \alphas (m_{\mathrm Z})
                               _{\overline{\mathrm{MS}}}                   $}}
\newcommand{\alphbar} {\mbox{$ \overline{\alpha}_{\mathrm s}               $}}
\newcommand{\Ptau}    {\mbox{$ P_{\tau}                                    $}}
\newcommand{\mean}[1] {\mbox{$ \left\langle #1 \right\rangle               $}}
\newcommand{\dgree}   {\mbox{$ ^\circ                                      $}}
% three particle states
\newcommand{\qqg}     {\mbox{$ {\mathrm q}\bar{\mathrm q}\gamma            $}}
\newcommand{\Wev}     {\mbox{$ W e \, \nu_e              $}}
\newcommand{\Zvv}     {\mbox{$ \Zn \nu \bar{\nu}                           $}}
\newcommand{\Zee}     {\mbox{$ Z^0 e^+ e^-                                 $}}
\newcommand{\ctw}     {\mbox{$ \cos\theta_{\mathrm W}                      $}}
\newcommand{\thw}     {\mbox{$ \theta_{\mathrm W}                          $}}
\newcommand{\thetabar}{\mbox{$ \theta^*                                    $}}
\newcommand{\phibar}  {\mbox{$ \phi^*                                      $}}
\newcommand{\thetapl} {\mbox{$ \theta_+                                    $}}
\newcommand{\phipl}   {\mbox{$ \phi_+                                      $}}
\newcommand{\thetamin}{\mbox{$ \theta_-                                    $}}
\newcommand{\phimin}  {\mbox{$ \phi_-                                      $}}
\newcommand{\ds}      {\mbox{$ {\mathrm d} \sigma                          $}}
\def    \ll           {\mbox{$\ell \ell                                    $}}
\def    \jjl          {\mbox{$j j \ell                           $}}
\def    \jj           {\mbox{$\jmath \jmath                                $}}
\def   \jjjj          {\mbox{${\it jets}                                   $}}
%  four particle states
\newcommand{\jjlv}    {\mbox{$ j j \ell \nu                                $}}
\newcommand{\jjvv}    {\mbox{$ j j \nu \bar{\nu}                           $}}
\newcommand{\qqvv}    {\mbox{$ \mathrm{q \bar{q}} \nu \bar{\nu}            $}}
\newcommand{\qqll}    {\mbox{$ \mathrm{q \bar{q}} \ell \bar{\ell}          $}}
\newcommand{\jjll}    {\mbox{$ j j \ell \bar{\ell}                         $}}
\newcommand{\lvlv}    {\mbox{$ \ell \nu \ell \nu                           $}}
\newcommand{\dz}      {\mbox{$ \delta g_{\mathrm{W W Z}    }               $}}
\newcommand{\pT}      {\mbox{$ p_{\mathrm{T}}                              $}}
\newcommand{\pt}      {\mbox{$ p_{\mathrm{t}}                              $}}
\newcommand{\ptr}     {\mbox{$ p_{\perp}                                   $}}
\newcommand{\ptrjet}  {\mbox{$ p_{\perp {\mathrm{jet}}}                    $}}
\newcommand{\Wvis}    {\mbox{$ {\mathrm W}_{\mathrm{vis}}                  $}}
\newcommand{\gamgam}  {\mbox{$ \gamma \gamma                               $}}
\newcommand{\qaqb}    {\mbox{$ {\, \mathrm q}_1 \,  \bar{\mathrm q}_2      $}}
\newcommand{\qcqd}    {\mbox{$ {\, \mathrm q}_3  \, \bar{\mathrm q}_4      $}}
\newcommand{\bbbar}   {\mbox{$ {\, \mathrm b} \, \bar{\mathrm b}           $}}
\newcommand{\ccbar}   {\mbox{$ {\, \mathrm c} \, \bar{\mathrm c}           $}}
\newcommand{\ffbar}   {\mbox{$ {\, \mathrm f} \, \bar{\mathrm f}           $}}
\newcommand{\ffbarp}  {\mbox{$ {\, \mathrm f} \, \bar{\mathrm f}'          $}}
\newcommand{\qqbar}   {\mbox{$\mathrm q \, \bar{\mathrm q}                 $}}
\newcommand{\quark}   {\mbox{$\mathrm q                                    $}}
\newcommand{\charm}   {\mbox{$\mathrm c                                    $}}
\newcommand{\bottom}  {\mbox{$\mathrm b                                    $}}
\newcommand{\topq}    {\mbox{$\mathrm t                                    $}}
\newcommand{\quarkb}   {\mbox{$\bar{\mathrm q}               $}}
\newcommand{\charmb}   {\mbox{$\bar{\mathrm c}               $}}
\newcommand{\bottomb}  {\mbox{$\bar{\mathrm b}               $}}
\newcommand{\topqb}    {\mbox{$\bar{\mathrm t}               $}}
\newcommand{\nunubar} {\mbox{$ {\, \nu} \, \bar{\nu}                       $}}
\newcommand{\qqbarp}  {\mbox{$ {\, \mathrm q'} \, \bar{\mathrm q}'         $}}
\newcommand{\djoin}   {\mbox{$ d_{\mathrm{join}}                           $}}
\newcommand{\mErad}   {\mbox{$ \left\langle E_{\mathrm{rad}} \right\rangle $}}
\newcommand{\bptre}{\rm b^{+}_{3}}
\newcommand{\bp}{\rm b^{+}_{1}}
\newcommand{\bo}{\rm b^0}
\newcommand{\bos}{\rm b^0_s}
\newcommand{\bss}{\rm b^s_s}
%    Meson decays
\newcommand{\BsDmX}{{B_{s}^{0}} \rightarrow D \mu X}
\newcommand{\BsDsm}{{B_{s}^{0}} \rightarrow D_{s} \mu X}
\newcommand{\BsDsX}{{B_{s}^{0}} \rightarrow D_{s} X}
\newcommand{\BDsX}{B \rightarrow D_{s} X}
\newcommand{\BDomX}{B \rightarrow D^{0} \mu X}
\newcommand{\BDpmX}{B \rightarrow D^{+} \mu X}
\newcommand{\Dsfmn}{D_{s} \rightarrow \phi \mu \nu}
\newcommand{\Dsfipi}{D_{s} \rightarrow \phi \pi}
\newcommand{\DsfX}{D_{s} \rightarrow \phi X}
\newcommand{\DpfX}{D^{+} \rightarrow \phi X}
\newcommand{\DofX}{D^{0} \rightarrow \phi X}
\newcommand{\DfX}{D \rightarrow \phi X}
\newcommand{\DsD}{B \rightarrow D_{s} D}
\newcommand{\DsmX}{D_{s} \rightarrow \mu X}
\newcommand{\DmX}{D \rightarrow \mu X}
\newcommand{\Zbb}{Z^{0} \rightarrow \rm b \overline{b}}
\newcommand{\Zcc}{Z^{0} \rightarrow \rm c \overline{c}}
\newcommand{\Rbb}{\frac{\Gamma_{Z^0 \rightarrow \rm b \overline{b}}}
{\Gamma_{Z^0 \rightarrow Hadrons}}}
\newcommand{\Rcc}{\frac{\Gamma_{Z^0 \rightarrow \rm c \overline{c}}}
{\Gamma_{Z^0 \rightarrow Hadrons}}}
\newcommand{\str}{\rm s \overline{s}}
%   mesons
\newcommand{\Bs}{\rm{B^0_s}}
\newcommand{\Bsb}{\overline{\rm{B^0_s}}}
\newcommand{\Bp}{\rm{B^{+}}}
\newcommand{\Bm}{\rm{B^{-}}}
\newcommand{\Bo}{\rm{B^{0}}}
\newcommand{\Bd}{\rm{B^{0}_{d}}}
\newcommand{\Bdb}{\overline{\rm{B^{0}_{d}}}}
\newcommand{\Lb}{\Lambda^0_b}
\newcommand{\Lbb}{\overline{\Lambda^0_b}}
\newcommand{\Kstar}{\rm{K^{\star 0}}}
\newcommand{\phim}{\rm{\phi}}
\newcommand{\Ds}{\mbox{D}_s}
\newcommand{\Dsp}{\mbox{D}_s^+}
\newcommand{\Dp}{\mbox{D}^+}
\newcommand{\Dn}{\mbox{D}^0}
\newcommand{\Dsb}{\overline{\mbox{D}_s}}
\newcommand{\Dm}{\mbox{D}^-}
\newcommand{\Dnb}{\overline{\mbox{D}^0}}
\newcommand{\Lc}{\Lambda_c}
\newcommand{\Lcb}{\overline{\Lambda_c}}
\newcommand{\Dstarp}{\mbox{D}^{\ast +}}
\newcommand{\Dstarm}{\mbox{D}^{\ast -}}
\newcommand{\Dsstarp}{\mbox{D}_s^{\ast +}}
\newcommand{\Pb}{P_{b-baryon}}
\newcommand{\KKpi}{\rm{ K K \pi }}
\newcommand{\nb}{\rm{nb}}
\newcommand{\Gm}{\rm{G_{\mu}}}
\newcommand{\Afb}{\rm{A_{FB}}}
\newcommand{\Afbs}{\rm{A_{FB}^{s}}}
\newcommand{\sigmaf}{\sigma_{\rm{F}}}
\newcommand{\sigmab}{\sigma_{\rm{B}}}
\newcommand{\NF}{\rm{N_{F}}}
\newcommand{\NB}{\rm{N_{B}}}
\newcommand{\Nnu}{\rm{N_{\nu}}}
\newcommand{\RZ}{\rm{R_Z}}
\newcommand{\rhob}{\rho_{eff}}
\newcommand{\Gammanz}{\rm{\Gamma_{Z}^{new}}}
\newcommand{\Gammani}{\rm{\Gamma_{inv}^{new}}}
\newcommand{\Gammasz}{\rm{\Gamma_{Z}^{SM}}}
\newcommand{\Gammasi}{\rm{\Gamma_{inv}^{SM}}}
\newcommand{\Gammaxz}{\rm{\Gamma_{Z}^{exp}}}
\newcommand{\Gammaxi}{\rm{\Gamma_{inv}^{exp}}}
\newcommand{\rhoZ}{\rho_{\rm Z}}
\newcommand{\swsq}{\sin^2\!\thw}
\newcommand{\swsqmsb}{\sin^2\!\theta_{\rm W}^{\overline{\rm MS}}}
\newcommand{\swsqbar}{\sin^2\!\overline{\theta}_{\rm W}}
\newcommand{\cwsqbar}{\cos^2\!\overline{\theta}_{\rm W}}
\newcommand{\swsqb}{\sin^2\!\theta^{eff}_{\rm W}}
\newcommand{\eeX}{{e^+e^-X}}
\newcommand{\gaga}{{\gamma\gamma}}
\newcommand{\eeg}{{e^+e^-\gamma}}
\newcommand{\mumug}{{\mu^+\mu^-\gamma}}
\newcommand{\qqb}{{q\bar{q}}}
\newcommand{\eegg}{e^+e^-\rightarrow \gamma\gamma}
\newcommand{\eeggg}{e^+e^-\rightarrow \gamma\gamma\gamma}
\newcommand{\eeee}{e^+e^-\rightarrow e^+e^-}
\newcommand{\eeeeee}{e^+e^-\rightarrow e^+e^-e^+e^-}
\newcommand{\eeeeg}{e^+e^-\rightarrow e^+e^-(\gamma)}
\newcommand{\eeeegg}{e^+e^-\rightarrow e^+e^-\gamma\gamma}
\newcommand{\eeeg}{e^+e^-\rightarrow (e^+)e^-\gamma}
\newcommand{\eemumu}{e^+e^-\rightarrow \mu^+\mu^-}
\newcommand{\eetautau}{e^+e^-\rightarrow \tau^+\tau^-}
\newcommand{\eehad}{e^+e^-\rightarrow {\rm hadrons}}
\newcommand{\eettg}{e^+e^-\rightarrow \tau^+\tau^-\gamma}
\newcommand{\eell}{e^+e^-\rightarrow l^+l^-}
\newcommand{\Ztopig}{{\rm Z}^0\rightarrow \pi^0\gamma}
\newcommand{\Ztogg}{{\rm Z}^0\rightarrow \gamma\gamma}
\newcommand{\Ztoee}{{\rm Z}^0\rightarrow e^+e^-}
\newcommand{\Ztoggg}{{\rm Z}^0\rightarrow \gamma\gamma\gamma}
\newcommand{\Ztomumu}{{\rm Z}^0\rightarrow \mu^+\mu^-}
\newcommand{\Ztotautau}{{\rm Z}^0\rightarrow \tau^+\tau^-}
\newcommand{\Ztoll}{{\rm Z}^0\rightarrow l^+l^-}
\newcommand{\Ztocc}{{\rm Z^0\rightarrow c \bar c}}
\newcommand{\Lamp}{\Lambda_{+}}
\newcommand{\Lamm}{\Lambda_{-}}
\newcommand{\Pt}{\rm P_{t}}
\newcommand{\Gee}{\Gamma_{ee}}
\newcommand{\Gpig}{\Gamma_{\pi^0\gamma}}
\newcommand{\Ggg}{\Gamma_{\gamma\gamma}}
\newcommand{\Gggg}{\Gamma_{\gamma\gamma\gamma}}
\newcommand{\Gmumu}{\Gamma_{\mu\mu}}
\newcommand{\Gtautau}{\Gamma_{\tau\tau}}
\newcommand{\Ginv}{\Gamma_{\rm inv}}
\newcommand{\Ghad}{\Gamma_{\rm had}}
\newcommand{\Gnu}{\Gamma_{\nu}}
\newcommand{\GnuSM}{\Gamma_{\nu}^{\rm SM}}
\newcommand{\Gll}{\Gamma_{l^+l^-}}
\newcommand{\Gff}{\Gamma_{f\overline{f}}}
\newcommand{\Gtot}{\Gamma_{\rm tot}}
\newcommand{\Rb}{\mbox{R}_b}
\newcommand{\Rc}{\mbox{R}_c}
\newcommand{\al}{a_l}
\newcommand{\vl}{v_l}
\newcommand{\af}{a_f}
\newcommand{\vf}{v_f}
\newcommand{\ael}{a_e}
\newcommand{\ve}{v_e}
\newcommand{\amu}{a_\mu}
\newcommand{\vmu}{v_\mu}
\newcommand{\atau}{a_\tau}
\newcommand{\vtau}{v_\tau}
\newcommand{\ahatl}{\hat{a}_l}
\newcommand{\vhatl}{\hat{v}_l}
\newcommand{\ahate}{\hat{a}_e}
\newcommand{\vhate}{\hat{v}_e}
\newcommand{\ahatmu}{\hat{a}_\mu}
\newcommand{\vhatmu}{\hat{v}_\mu}
\newcommand{\ahattau}{\hat{a}_\tau}
\newcommand{\vhattau}{\hat{v}_\tau}
\newcommand{\vtildel}{\tilde{\rm v}_l}
\newcommand{\avsq}{\ahatl^2\vhatl^2}
\newcommand{\Ahatl}{\hat{A}_l}
\newcommand{\Vhatl}{\hat{V}_l}
\newcommand{\Afer}{A_f}
\newcommand{\Ael}{A_e}
\newcommand{\Aferb}{\bar{A_f}}
\newcommand{\Aelb}{\bar{A_e}}
\newcommand{\AVsq}{\Ahatl^2\Vhatl^2}
\newcommand{\Iwk}{I_{3l}}
\newcommand{\Qch}{|Q_{l}|}
\newcommand{\roots}{\sqrt{s}}
\newcommand{\mt}{m_t}
\newcommand{\Rechi}{{\rm Re} \left\{ \chi (s) \right\}}
\newcommand{\up}{^}
\newcommand{\abscosthe}{|cos\theta|}
\newcommand{\dsum}{\Sigma |d_\circ|}
\newcommand{\zsum}{\Sigma z_\circ}
\newcommand{\sint}{\mbox{$\sin\theta$}}
\newcommand{\cost}{\mbox{$\cos\theta$}}
\newcommand{\mcost}{|\cos\theta|}
\newcommand{\epair}{\mbox{$e^{+}e^{-}$}}
\newcommand{\mupair}{\mbox{$\mu^{+}\mu^{-}$}}
\newcommand{\taupair}{\mbox{$\tau^{+}\tau^{-}$}}
\newcommand{\fullskip}{\vskip 16cm}
\newcommand{\halfskip}{\vskip  8cm}
\newcommand{\quarskip}{\vskip  6cm}
\newcommand{\abitskip}{\vskip 0.5cm}
\newcommand{\ba}{\begin{array}}
\newcommand{\ea}{\end{array}}
\newcommand{\bc}{\begin{center}}
\newcommand{\ec}{\end{center}}
\newcommand{\be}{\begin{eqnarray}}
\newcommand{\eeq}{\end{eqnarray}}
\newcommand{\bes}{\begin{eqnarray*}}
\newcommand{\ees}{\end{eqnarray*}}
\newcommand{\Kz}{\ifmmode {\rm K^0_s} \else ${\rm K^0_s} $ \fi}
\newcommand{\Zz}{\ifmmode {\rm Z^0} \else ${\rm Z^0 } $ \fi}
\newcommand{\xxbar}{\ifmmode {\rm x\bar{x}} \else ${\rm x\bar{x}} $ \fi}
\newcommand{\rphi}{\ifmmode {\rm R\phi} \else ${\rm R\phi} $ \fi}
%%%%%%%%%%%%%%%%%%%%%%%
% End of Declarations S.K %
%%%%%%%%%%%%%%%%%%%%%%%

\newcommand{\Lum}{${\cal L}\;$}
\newcommand{\lum}{{\cal L}}
\newcommand{\Cms}{$\mbox{ cm}^{-2} \mbox{ s}^{-1}\;$}
\newcommand{\cms}{\mbox{ cm}^{-2} \mbox{ s}^{-1}\;}
\newcommand{\Ecms}    {\mbox{$ E_{\mathrm{\small cms}}                      $}}
\newcommand{\Ecm}    {\mbox{$ E_{\mathrm{\small cm}}                      $}}
\newcommand{\Etvis}    {\mbox{$ E^{T}_{\mathrm{\small vis}}                      $}}
\newcommand{\Evis}    {\mbox{$ E_{\mathrm{\small vis}}                      $}}
\newcommand{\Erad}    {\mbox{$ E_{\mathrm{\small rad}}                      $}}
\newcommand{\Mvis}    {\mbox{$ M_{\mathrm{\small vis}}                      $}}
\newcommand{\pvis}    {\mbox{$ p_{\mathrm{\small vis}}                      $}}
\newcommand{\Minv}    {\mbox{$ M_{\mathrm{\small inv}}                      $}}
\newcommand{\pmiss}   {\mbox{$ p_{\mathrm{\small miss}}                     $}}
\newcommand{\Mhfit}{\; \hat{m}_{H^0} }
\newcommand{\bl}      {\mbox{\ \ \ \ \ \ \ \ \ \ } }
%%%%%%%%%%%%%%%%%%%%%%%
% End of Declarations J.M %
%%%%%%%%%%%%%%%%%%%%%%%
\newcommand{\Zto}   {\mbox{$\mathrm Z^0 \to$}}
\newcommand{\etal}  {\mbox{\it et al.}}
\def\NPB#1#2#3{{\rm Nucl.~Phys.} {\bf{B#1}} (#2) #3}
\def\PLB#1#2#3{{\rm Phys.~Lett.} {\bf{B#1}} (#2) #3}
\def\PRD#1#2#3{{\rm Phys.~Rev.} {\bf{D#1}} (#2) #3}
\def\PRL#1#2#3{{\rm Phys.~Rev.~Lett.} {\bf{#1}} (#2) #3}
\def\ZPC#1#2#3{{\rm Z.~Phys.} {\bf C#1} (#2) #3}
\def\PTP#1#2#3{{\rm Prog.~Theor.~Phys.} {\bf#1}  (#2) #3}
\def\MPL#1#2#3{{\rm Mod.~Phys.~Lett.} {\bf#1} (#2) #3}
\def\PR#1#2#3{{\rm Phys.~Rep.} {\bf#1} (#2) #3}
\def\RMP#1#2#3{{\rm Rev.~Mod.~Phys.} {\bf#1} (#2) #3}
\def\HPA#1#2#3{{\rm Helv.~Phys.~Acta} {\bf#1} (#2) #3}
\def\NIMA#1#2#3{{\rm Nucl.~Instr.~and~Meth.} {\bf#1} (#2) #3} 
\def\CPC#1#2#3{{\rm Comp.~Phys.~Comm.} {\bf#1} (#2) #3}
\def\EPJC#1#2#3{{\rm E.~Phys.~J.} {\bf{C#1}} (#2) #3}
% Imported from chargino paper
\def    \DM          {\mbox{$\Delta$M}}
\def    \missEt      {\ifmmode{/\mkern-11mu E_t}\else{${/\mkern-11mu E_t}$}\fi}
\def    \missE       {\ifmmode{/\mkern-11mu E}\else{${/\mkern-11mu E}$}\fi}
\def    \missp       {\ifmmode{/\mkern-11mu p}\else{${/\mkern-11mu p}$}\fi}
\def    \misspt      {\ifmmode{/\mkern-11mu p_t}\else{${/\mkern-11mu p_t}$}\fi}
\def    \DML         {\mbox{5~GeV $<\Delta M<$ 10~GeV}}
\def    \rs          {\mbox{$\sqrt{s}$}}
\def    \msneu       {\mbox{$m_{\tilde{\nu}}$}}
\def    \rad         {\mbox{$\it{rad}$}}

\title{Summary of ILD performance at SPS1a'}
%%\LaTeX Template for LCWS08/ILC08 Proceedings} %% 
%***********************************************************************
% AUTHORS INFORMATION AREA
%***********************************************************************
\author{Mikael Berggren,
Nicola d'Ascenzo,
Peter Schade, and
Olga Stempel
%--- \author{First Author$^1$ and Second Author$^2$
% Optional short acknowledgment: remove next line if non-needed
\thanks{Support by DFG through SFB 676 is acknowledged}
% DO NOT MODIFY THE FOLLOWING '\vspace' ARGUMENT
\vspace{.3cm}\\
% Addresses and institutions (remove "1- " in case of a single institution)
 DESY -FLC \\ 
Notekstr. 85, 22607 Hamburg -Germany 
}
%%***********************************************************************
% END OF AUTHORS INFORMATION AREA
%***********************************************************************

\maketitle

\begin{abstract}
The performance of the ILD detector at the ILC for 
the analysis of $\mu$ and $\tau$ channels at the
SUSY benchmark-point SPS1a' has been studied with full
detector simulation. It is concluded that
if 500 fb$^{-1}$ is delivered to the experiment,
$\Delta(\MXN{1}) = 920\ \MeVcc$,
$\Delta(\msmul) = 100\ \MeVcc$, $\Delta(\MXN{2}) = 1.38 ~\GeVcc$,
and $\Delta(\sigma(\eeto \smul \smul))$ = 1.35 fb can
be achieved from the $\mu$ channels alone.
The preliminary results from the $\stau$ channels,
indicates that $\Delta(P_{\tau}) = 13 $ \% is also
achievable.
\end{abstract}

\section{Introduction}
The SUSY benchmark point SPS1a'~\cite{bib:sps1ap} offers a rich
phenomenology at the ILC. It is point with quite
low mass-spectrum in the slepton sector, and heavy
squarks. Bosinos up to $\XN{3}$ (in  $\eeto \XN{1} \XN{3}$)
would be produced at $E_{CMS}$ = $500 ~\GeV$. 
It is a pure mSUGRA model, hence R-parity and CP is conserved.  
The unification scale parameters are:
$M_{1/2} = 250 ~~\GeV , M_0 = 70 ~\GeV , A_0 = -300 ~\GeV ,
\tan{\beta} = 10$, and  $sign(\mu) = +1 $.
The point is not in contradiction with any experimental limits~\cite{bib:pmssm}.
The $\stau$ is the NLSP, and $\mstone = 107.9 ~\GeVcc$ and  
$\MXN{1} = 97.7 ~\GeVcc$, so $\Delta(M) = 10.2 ~\GeVcc$.
At E$_{CMS}$ = $500 ~\GeV$, this yields
$P_{\stau,min} = 2.2 ~\GeVc$
% , P_{\stau,max} = 42.8 ~\GeVc$,
hence $\gamma \gamma$ events will pose a problem.
As SPS1a' is a point with an important co-annihilation contribution
to the dark-matter relic density, the $\mstone$ is a most
important quantity to determine.
An other consequence of the $\stau$ being the NLSP, is that $\tau$:s 
are present in large fraction of the SUSY decays, so that
SUSY itself will be a mayor background source for $\tau$ channels.
On the other hand, the $\msmul$($\msmur$) is 189.9(125.3) $~\GeVcc$,
so that the minimum $\mu$ energy is 32.1(6.6) $~\GeV$. As, in addition, the
branching ratios to $\smu$ in bosino decays are quite low, the $\mu$
final states offer cleaner conditions, and are well suited for
doing the most precise measurements.

The present note reports on the status of the analysis of
of the $\mu$ and $\tau$ channels of the SPS1a' scenario in the LDC' detector.
\texttt{SPheno}~\cite{bib:spheno} was used to run the unification-scale
model to the EW scale, and \texttt{Whizard}~\cite{bib:Whizard} was
the used to generate events.
The \texttt{LDCPrime\_02Sc} detector model was
fully simulated using \texttt{MOKKA}\cite{mokka},
and the events were
reconstructed with \texttt{MarlinReco}~\cite{bib:MarlinReco}.
The same chain was used to produce background events.

\section{Analysis of $\mu$ channels}
Two channels containing only $\mu$:s in the final state was
chosen as a first study ~\cite{ndascthesis}:
$\smul \smul \rightarrow \mu \mu \XN{1} \XN{1} $
and $\XN{1} \XN{2} \rightarrow \mu \smur \XN{1}   \rightarrow \mu \mu \XN{1} \XN{1} $.  
\begin{figure}[h]
\centerline{
       \includegraphics[width=0.45\columnwidth]{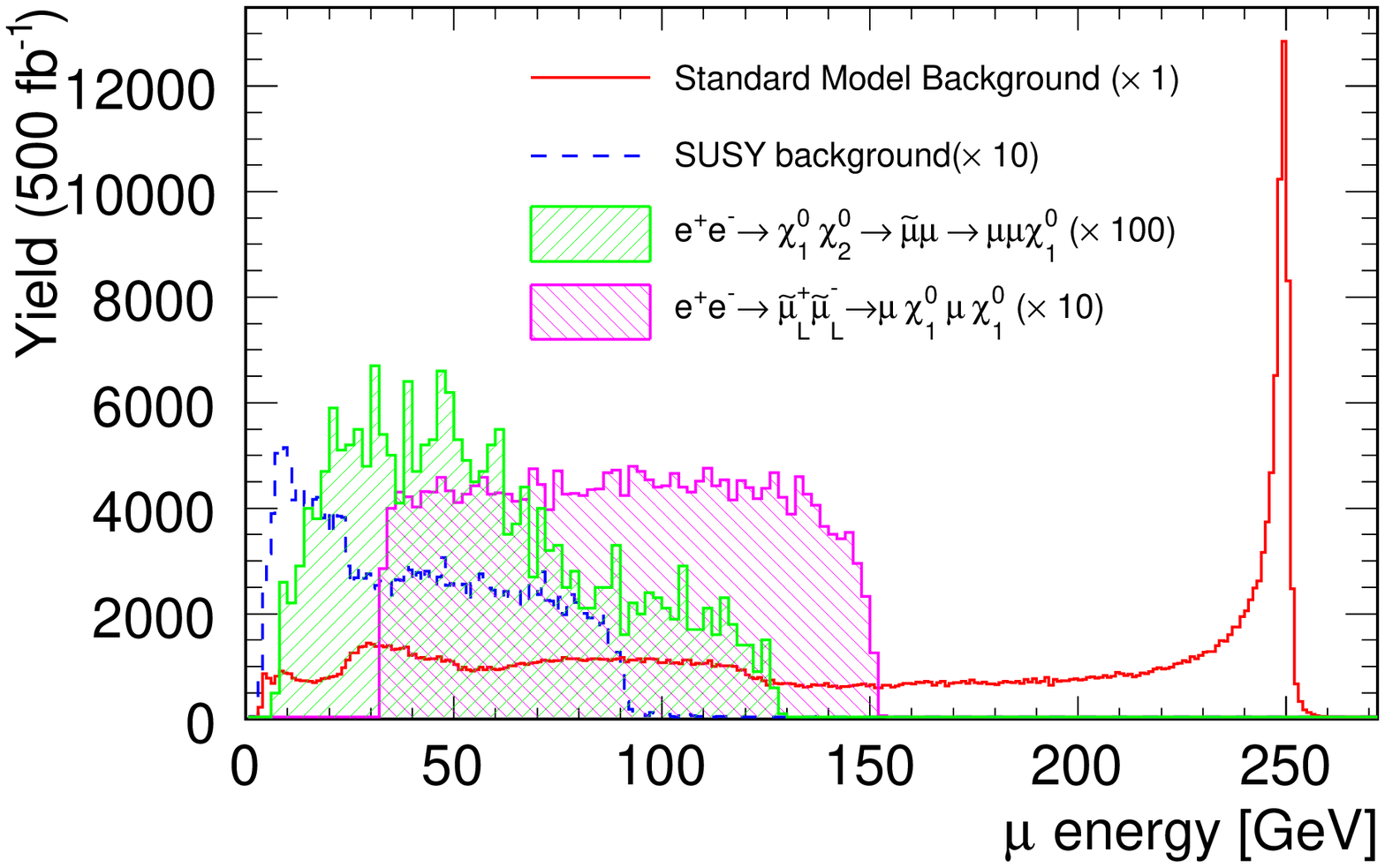}
       \includegraphics[width=0.45\columnwidth]{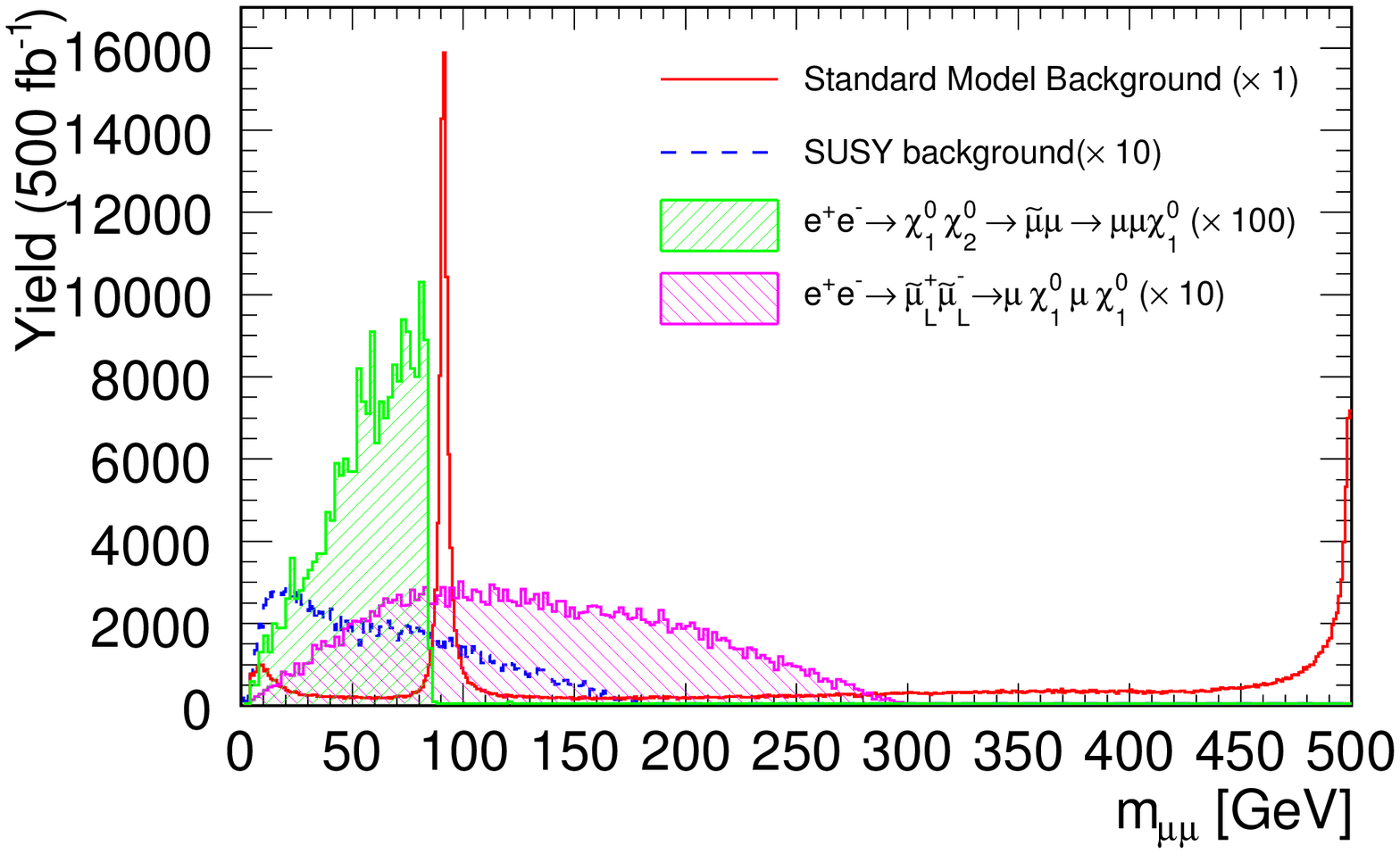}
}
\caption{The distributions of $P_{\mu}$ (left), and
M$_{\mu\mu}$ (right) }\label{fig:pmunmmumu}
\end{figure}
As mentioned in the introduction, the SUSY background problem
is not too severe in the $\mu$ channels, and it is advantageous to
run the ILC at the polarisation giving the largest signal.
Hence, these channels were studied assuming
80 \% left  e$^-$ polarisation and 60 \% right e$^+$ polarisation.
Under these conditions, the $\smul \smul$  process has a large cross-section, 
and is well suited to determine $\msmul$ and $\MXN{1}$.
$\XN{1} \XN{2}$ has a small cross-section $\times$ BR, but can be
used to determine $\MXN{2}$, without the need to scan over the
threshold.
The main background processes are  
other SUSY giving two $\mu$:s, mainly $\smur \smur$,
$\XN{2} \XN{2}$ with one $\XN{2}$ going to $\smu \mu$, the
other to $\snu \nu$
and
$\stau \stau$ with $\tau \rightarrow \mu \nu_{\mu} \nu_{\tau}$
Standard model background is mainly from $WW$ and $ZZ$.
Finally,
each of the two processes is background to the other one.

\begin{wrapfigure}{r}{0.5\columnwidth}
\centerline{
       \includegraphics[scale=0.35]{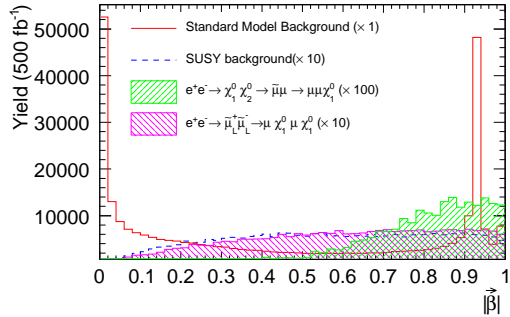}
}
\caption{The distribution of the velocity $\beta$}\label{fig:beta}
\end{wrapfigure}

The following kinematic variables were used to disentangle
signal and background, and to separate the two signal channels:
The momentum of $\mu$:s ($P_{\mu}$), the acolinearity angle between
the $\mu$:s ($\theta_{acol}$), the acoplanarity angle between them,defined as
the acolinearity in the projection perpendicular to the
beam-axis ($\theta_{acop}$),
the total missing transverse momentum ($P_{T miss}$) in the event, 
the invariant mass
of the two $\mu$:s (M$_{\mu\mu}$), the total
missing energy (E$_{miss}$), the polar angle of the missing
momentum ($\theta_{missing\ p}$), and the velocity $\beta$ of 
the $\mu$ system.
The distributions of $P_{\mu}$ and M$_{\mu\mu}$ are shown
in Fig.~\ref{fig:pmunmmumu}, and that of $\beta$ in Fig.~\ref{fig:beta}

\begin{figure}[h]
\centerline{
       \includegraphics[width=0.45\columnwidth]{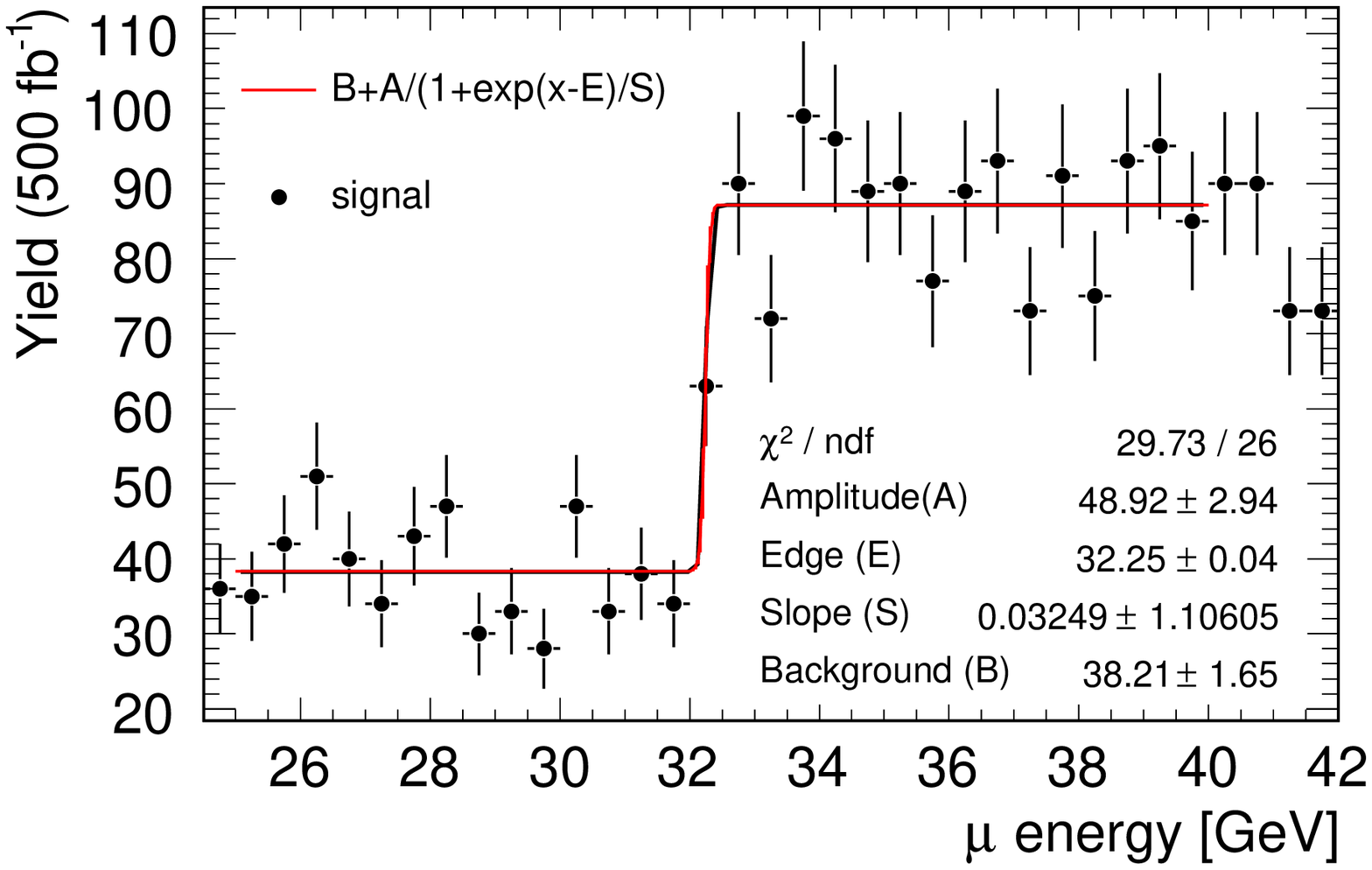}
       \includegraphics[width=0.45\columnwidth]{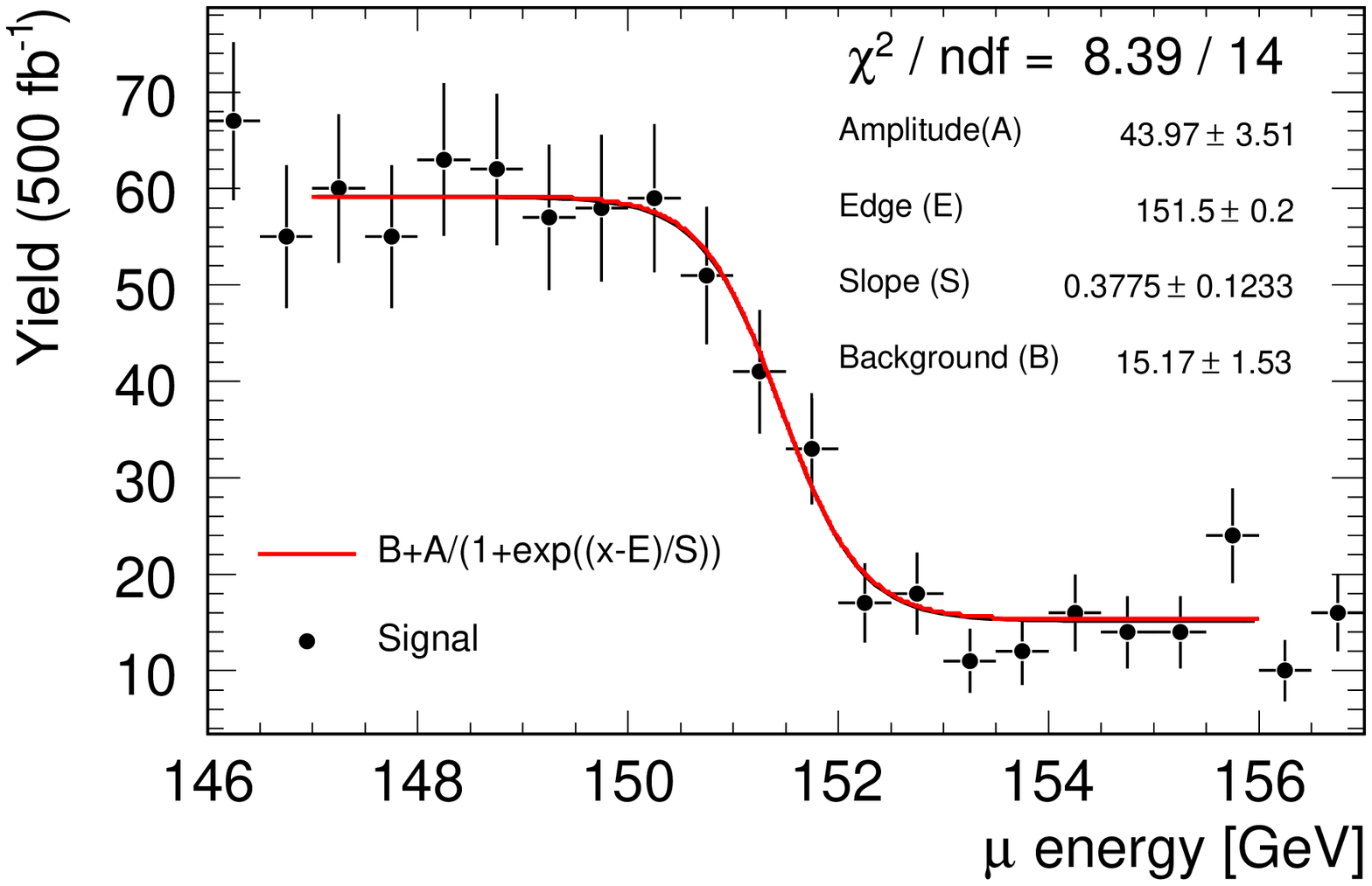}
}
\caption{The fit to the lower (left) and upper (right) edges
of the $P_{\mu}$ distribution }\label{fig:boxfit}
\end{figure}

The $\smul \smul$ channel was selected by demanding that
E$_{miss} \in [200,430] ~\GeV $,
M$_{\mu\mu} \notin [80,100] ~\GeV  $ and $ < 30   ~\GeVcc$,
and $\theta_{missing\ p}  \in [0.1 \pi , 0.9 \pi ]$.
Assuming an integrated luminosity of 500 fb$^{-1}$,
this leaves 13000 events of SM background and 11000 events
of SUSY background, while 16300 signal events were selected,
corresponding to an efficiency of 60 \%.
The $\smul$ and $\XN{1}$ masses were then extracted by
fitting the edges of the $P_{\mu}$ distribution, see Fig.~\ref{fig:boxfit}.
The errors on the fitted masses are $\Delta(\msmul) = 100 ~\MeVcc$ and         
$\Delta(\MXN{1}) = 920 ~\MeVcc$, respectively.
The beam-energy spread dominates these numbers.
The production cross-section was determined using the extended likelihood 
formed by $L(p_{T \mu}, \theta_{acol}))$, as these two variables were
not used in selecting the signal. The uncertainty on the
observed value is $\Delta(\sigma(\eeto \smul \smul))$ = 1.35 fb.

The $\XN{1} \XN{2}$ channel was selected by demanding that
$\theta_{missing\ p}  \in [0.2 \pi , 0.8 \pi ]$,
$\beta > 0.6$,
E$_{miss} \in [355,395] ~\GeV $,
p$_{T miss} > 40 ~\GeVc $, 
M$_{\mu\mu} \in [40, 85] ~\GeVcc$
and
E$_{viss} > 40 ~\GeVcc$.

\begin{wrapfigure}{r}{0.5\columnwidth}
\centerline{
       \includegraphics[width=0.45\columnwidth]{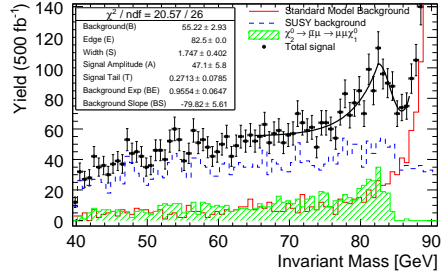}
}
\caption{The fit to the M$_{\mu\mu}$ distribution}\label{fig:edgefit}
\end{wrapfigure}

At the assumed integrated luminosity,
500 events of SM background and 2400 events
of SUSY background is expected, while 720 signal events were selected,
corresponding to an efficiency of 34 \%.
Assuming that the $\XN{1}$ mass is known from the
previous channel, the mass of $\XN{2}$ can be extracted
by a fit to the edge in the invariant mass spectrum, Fig.~\ref{fig:edgefit},
and an uncertainty of  
$\Delta(\MXN{2}) = 1.38 ~\GeVcc$ was found.

\section{$\tau$ channels}

As mentioned in the introduction,
SUSY itself poses a background problem in the $\tau$ analysis,
and it is therefore needed to run the ILC at the
polarisation that minimises the background.
For
100 \% left  e$^-$ polarisation and 100 \% right e$^+$ polarisation,
the cross-sections for $\XN{2} \XN{2}$ and $\XP{1} \XM{1}$ are
several hundred fb, and the branching ratios to $\stau$ is above
50 \%. With the opposite polarisation, however, these cross-sections
will almost vanish.
Hence, these channels were studied assuming
80 \% right  e$^-$ polarisation and 60 \% left e$^+$ polarisation.

As the $\gamma \gamma$ background poses another challenge
for the $\tau$ channels, quite strong criteria must be applied:
A correlated cut in $\rho$ (the transverse momentum of the jets
wrt. the thrust axis, in the projection perpendicular to the beam)
was also done: $\rho > 3\sin{\theta_{acop}}+1.7$.
To further reduce the $\gamma\gamma$ background to
acceptable levels, it was demanded that there be no significant
activity in the BeamCal, and that the $\phi$ angle of the
missing momentum was not in the direction of the
incoming beam-pipe.

The $\stau$ mass can be extracted from the end-point of the 
spectrum of E$_{\tau}$, which is equal to  $E_{\stau,max}$, 
and the $\XN{1}$ mass, known eg. from the
$\smul$ analysis above. In principle, the maximum of
the spectrum spectrum is at $P_{\stau,min}$, so that the $\stau$ can
be used to find $\MXN{1}$ as well, but due to the large
$\gamma \gamma$ background, the maximum is quite hard
to observe. 
     
\begin{wrapfigure}{r}{0.5\columnwidth}
\centerline{
       \includegraphics[width=0.41\columnwidth]{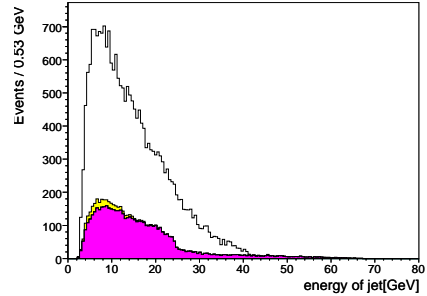}
}
\caption{E$_{jet}$ distribution after all cuts in the
decay-mode independent analysis}\label{fig:jetspect}
\end{wrapfigure}
To extract the signal in order to determine the end-point the
following cuts were applied:
$E_{miss} \in [430,490] ~\GeV$, 
$M_{jet} < 2 ~\GeVcc$,
$\theta_{jet}$ above 20 degrees,
$\theta_{acop} < 160$ degrees,
$\theta_{acol} \in [80,170]$ degrees,
$|\cos{\theta_{missing p}}| < 0.9$, and
charge of each jet = $\pm 1$.
In addition, the anti-$\gamma\gamma$ cut described above was applied.
After these cuts, the SM background was 222 events, the SUSY
background was 2747, while 8262 signal events remained (10.2 \%
efficiency). Fig.~\ref{fig:jetspect} shows that the end-point is
almost background free, and also that the turn-over point
(expected to be at $P_{\stau,min} = 2.2 ~\GeVc$)
is too distorted by the cuts to be useful.

\begin{wrapfigure}{r}{0.5\columnwidth}
\centerline{
         \includegraphics[width=0.41\columnwidth]{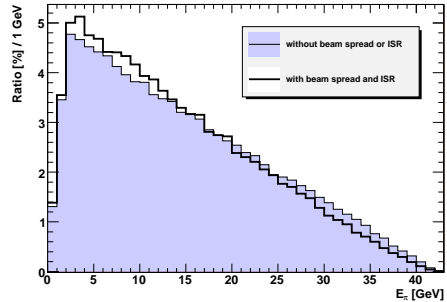}
}
\caption{The $P_\pi$ distribution, at generator level. Shaded:
fixed beam-energy, open: ISR and beam-spread included}\label{fig:truepispect}
\end{wrapfigure}

The $\stau$ mass-eigenstates are expected to be different
from the chiral ones, and the off-diagonal term of mass-matrix
is
$-M_{\tau}(A_{\stau} - \mu \tan{\beta})$. 
The diagonal
terms in the mass matrix are known from $\msmul$ and $\msmur$, so
a measurement of $\theta_{mix}$ gives $A_{\stau} - \mu \tan{\beta}$.
If $\XN{1}$ is purely bino - it is in SPS1a' - the $\tau$ polarisation 
(P$_{\tau}$) depends only on $\theta_{mix}$. 
P$_{\tau}$ can be extracted from spectrum for exclusive 
decay-mode(s). In this analysis, the $\tau \rightarrow \pi^{+-} \nu_{\tau}$
mode has been studied.
The spectrum of $\pi$:s in  the decay-chain $\stau \rightarrow \tau 
\rightarrow \pi^{+-} \nu_{\tau}$ is shown in Fig.~\ref{fig:truepispect},
with and without ISR and beam-spread.
The highest sensitivity to the polarisation is
in the region with  $P_\pi < P_{\stau,min}$.

The  $\stau \rightarrow \tau \rightarrow \pi^{+-} \nu_{\tau}$ signal
is selected with a set of cuts that intend to distort the
spectrum as little as possible. The following pre-selections
were first applied:
The events should pass the anti-$\gamma\gamma$ cut, 
$E_{vis}$ should be  $< 120 ~\GeV$,
the number of reconstructed particles $<$ 20,
and at least one of the two jets should contain a single particle.
This single particle should be identified as a $\pi$, and
have $E < 43 ~\GeV$. Finally, the total charge should be 0.
Events passing this preslection should
then also fulfil the following criteria:
The mass of the rest of the event (ie. after removing the
signal pion) should be below 2.5 $~\GeVcc$ 
$|\cos{\theta}|$ of both jets should be $< 0.9$, and
$\theta_{acop}$ should be above 85 degrees.
Finally, the sum over the two jets of the $p_T$ of one jet wrt. the 
direction of the
other should be below 30 $~\GeVc$.

\begin{figure}[h]
%{r}{0.5\columnwidth}
\centerline{
       \includegraphics[width=0.41\columnwidth]{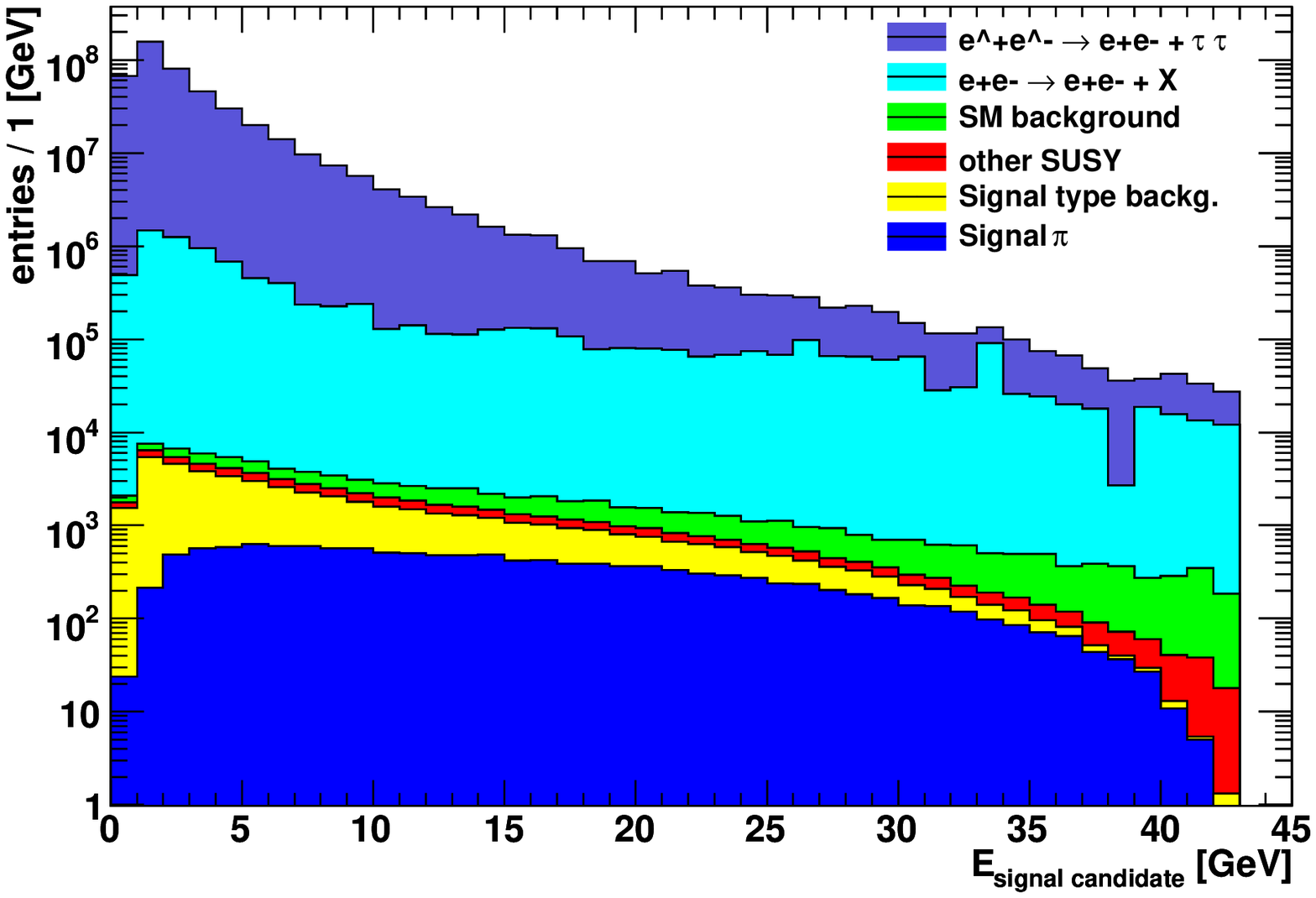}
       \includegraphics[width=0.41\columnwidth]{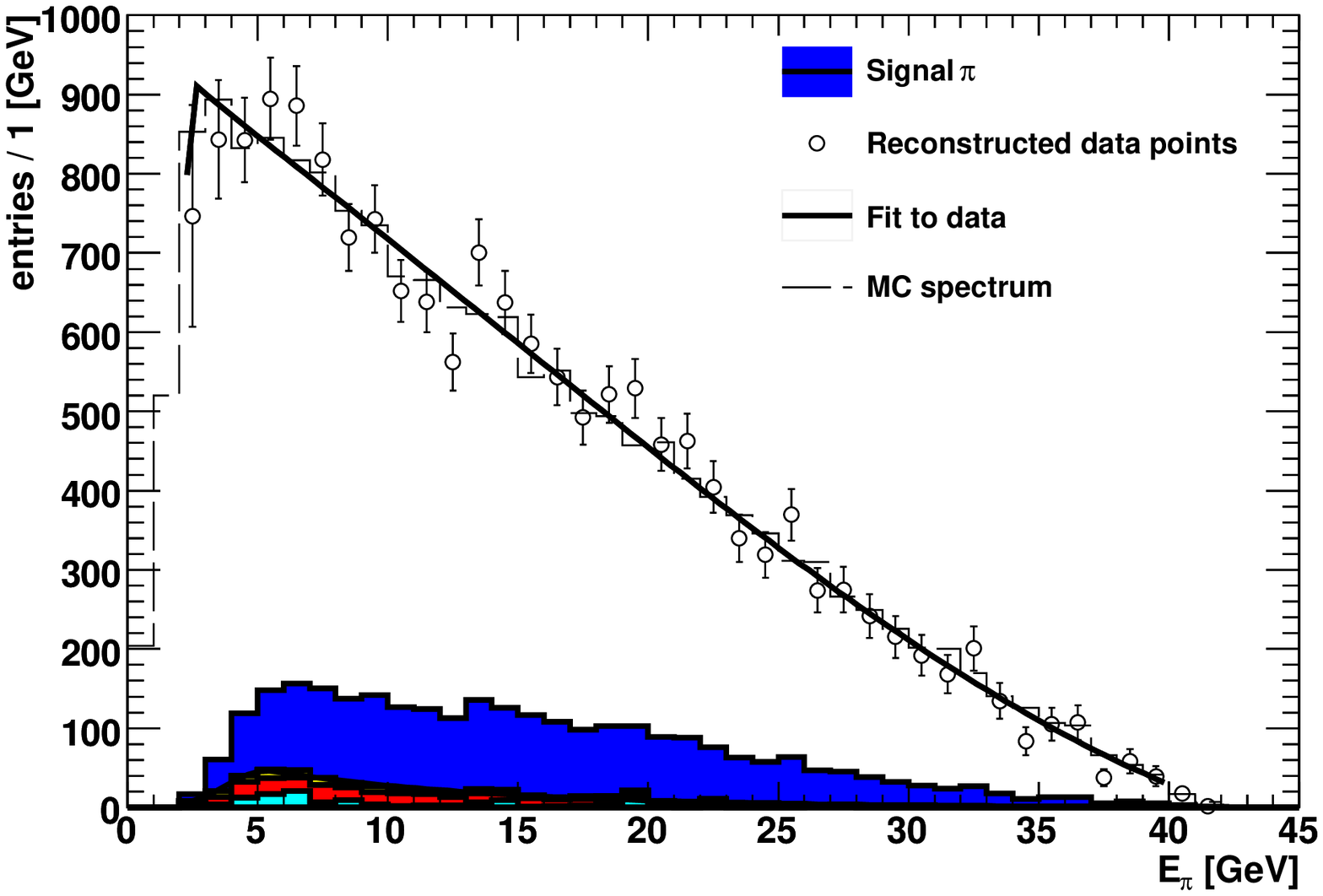}
}
\caption{Distribution of $P_\pi$ before (left) and after (right) cuts.
The right-hand plot also shows the spectrum after background-subtraction and
efficiency correction (dots), and the final fit.}\label{fig:pispect}
\end{figure}

With these cuts, 134 SM jets remain, and 373 SUSY jets,
while 2311 signal-jets are retained (13 \%).
The initial and final $\pi$ spectra are shown in  Fig.~\ref{fig:pispect}.
The procedure to extract the polarisation in the presence of background
is to first fit the simulated background alone to a heuristic function.
\footnote{When real data is available, the simulation of the background
can be verified by reversing cuts to select a signal-free, but
SUSY-dominated region in the parameter-space.}
The signal
selections cuts are then applied to the signal+background
sample, 
and the function is subtracted from the observed distribution.
An efficiency correction function, determined from signal-only
simulation, is applied. The resulting distribution is then fitted
with the theoretical spectrum, corrected for ISR and beam-spread,
and the polarisation is obtained, see  Fig.~\ref{fig:pispect}, right.
Assuming an integrated luminosity of 500 fb$^{-1}$,
the value found is  P$_{\tau} = (93 \pm 13)$ \%,
where the error also includes the uncertainty of the
background parametrisation.

\section{Conclusions}

A study of some channels in SPS1a' SUSY scenario fully
simulated in the LDC' detector at the ILC was presented.
By analysing the channel $\eeto \smul \smul$,  
it was concluded that $\Delta(\MXN{1}) = 920 \MeVcc$,
$\Delta(\msmul) = 100 \MeVcc$ and
$\Delta(\sigma(\eeto \smul \smul))$ = 1.35 fb, could be
attained with an integrated luminosity of 500 fb$^{-1}$ with 
80 \% left  e$^-$ polarisation and 60 \% right e$^+$ polarisation.
In the channel $\XN{1} \XN{2} \rightarrow \mu \smur \XN{1}   
\rightarrow \mu \mu \XN{1} \XN{1} $, $\Delta(\MXN{2}) = 1.38 ~\GeVcc$,
was found, under the same conditions. It should be noted that 
this value is comparable to what a dedicated scan of the
$\XN{2} \XN{2}$ threshold would give.

In addition, a progress report on $\stau$ production was
given. The preliminary result on the measurement
of the $\tau$ polarisation gives $\Delta($P$_{\tau}) = 13 $ \%.
Note, however, that this requires that the beam-polarisations are
{\it opposite} to what the was used in the $\mu$ channel.

% ****************************************************************************
% BIBLIOGRAPHY AREA
% ****************************************************************************

\begin{footnotesize}
% IF YOU DO NOT USE BIBTEX, USE THE FOLLOWING SAMPLE SCHEME FOR THE REFERENCES
% ----------------------------------------------------------------------------

\end{footnotesize}

% ****************************************************************************
% END OF BIBLIOGRAPHY AREA
% ****************************************************************************

\end{document}